\documentclass[conference]{IEEEtran}
\IEEEoverridecommandlockouts

\usepackage{cite}
\usepackage{amsmath,amssymb,amsfonts}
\usepackage{algorithmic}
\usepackage{graphicx}
\usepackage{textcomp}

\usepackage[inline]{enumitem}
\usepackage{algorithmic}
\usepackage{graphicx}
\usepackage{textcomp}
\usepackage{xcolor}

\usepackage{witharrows}

\usepackage[flushleft]{threeparttable}
\usepackage{tablefootnote}
\usepackage{array}

\newcolumntype{P}[1]{>{\centering\arraybackslash}m{#1}}
\usepackage{multicol}
\usepackage{multirow}
\usepackage{tabulary}
\usepackage{colortbl}
\definecolor{mygray}{gray}{0.90}
\usepackage{makecell}
\usepackage{booktabs}
\usepackage{subcaption}
\usepackage{stmaryrd}
\usepackage{float}
\usepackage{pifont}

\usepackage{arydshln}
\colorlet{mygray}{gray!15!white}

\usepackage{url,hyperref,microtype}
\hypersetup{
    colorlinks=true,
    citecolor=blue,
    linkcolor=blue,
    filecolor=magenta,      
    urlcolor=black,
}

\def\BibTeX{{\rm B\kern-.05em{\sc i\kern-.025em b}\kern-.08em
    T\kern-.1667em\lower.7ex\hbox{E}\kern-.125emX}}
\begin{document}

\title{One-Block Transformer (1BT) for EEG-Based Cognitive Workload Assessment
}

\author{\IEEEauthorblockN{Stefanos Gkikas}
\IEEEauthorblockA{\textit{Honda Research Institute Japan} \\
Wako City, Japan \\
stefanos.gkikas@jp.honda-ri.com}
\and
\IEEEauthorblockN{Christian Arzate Cruz}
\IEEEauthorblockA{\textit{Honda Research Institute Japan} \\
Wako City, Japan  \\
christian.arzate@jp.honda-ri.com}
\and
\IEEEauthorblockN{Thomas Kassiotis}
\IEEEauthorblockA{\textit{Department of Electronic Engineering} \\
\textit{Hellenic Mediterranean University}\\
Chania, Greece \\
ddk305@edu.hmu.gr}
\and
\IEEEauthorblockN{Giorgos Giannakakis}
\IEEEauthorblockA{\textit{Department of Electronic Engineering} \\
\textit{Hellenic Mediterranean University}\\
Chania, Greece \\
ggian@hmu.gr}
\and
\IEEEauthorblockN{Raul Fernandez Rojas}
\IEEEauthorblockA{\textit{BioSIS (Biosensing \& Intelligent Systems) Lab}\\ 
\textit{Centre for Intelligent Computing and Systems} \\
\textit{University of Canberra}\\
Canberra, Australia\\
raul.fernandezrojas@canberra.edu.au}
\and
\IEEEauthorblockN{Randy Gomez}
\IEEEauthorblockA{\textit{Honda Research Institute Japan} \\
Wako City, Japan \\
r.gomez@jp.honda-ri.com}
}

\maketitle

\begin{abstract}

Accurate and continuous estimation of cognitive workload is fundamental to creating adaptive human--machine systems. However, designing architectures that balance representational capacity with computational efficiency has been challenging for practical deployment. This paper introduces \textit{1BT}, a One-Block Transformer for compact and efficient EEG-based cognitive workload assessment. The model aggregates multi-channel temporal sequences via a minimal latent bottleneck, using a single cross-attention module followed by lightweight self-attention. A controlled study involving 11 participants performing three cognitively diverse tasks (abstract reasoning, numerical problem-solving, and an interactive video game) was conducted with continuous EEG recordings across two workload levels.
Systematic architectural analysis identifies the most compact configuration that preserves high performance, while substantially lowering computational cost. The final model achieves high workload classification performance with under $0.5$ million parameters and $0.02$ GFLOPs, paving the way for a design direction for real-time cognitive workload monitoring in resource-constrained settings.

\end{abstract}


\section{Introduction}
Cognitive load denotes the mental effort required by working memory during task performance and lies at the core of cognitive and affective computing. The capacity to estimate cognitive load in real time has grown increasingly important for human--machine interaction, particularly in high-stakes domains such as training and education systems, transportation, automation, robotics~\cite{arzate_3dmm_2022}, and aerospace. In these contexts, adaptive and responsive support can meaningfully enhance performance, efficiency, and safety \cite{clare_2025, fridman_reimer_2018}.
Beyond specialized environments, cognitive load arises naturally across the full spectrum of daily life, spanning professional roles in healthcare, aviation, education, and other high-performance settings \cite{anders_moontaha_2024, hemakom_atiwiwat_2024}, as well as routine personal responsibilities such as household management and childcare \cite{reichstiebert_froehlich_2023}.
Modeling these internal states relates to broader work in information-based behavioral analysis~\cite{chatziadam_dimitraidis_2020, 10807220}, where similar computational challenges arise.
Recognizing how cognitive load changes across these diverse contexts is critical for developing systems that provide context-aware assistance and help users sustain effective performance over time. When elevated demands persist, they often overlap with stress-related processes, making continuous, accurate, and unobtrusive workload monitoring a practical priority \cite{hassard_teoh_2018}. These challenges are not unique to cognitive workload; related estimation problems appear in pain assessment and other physiological state recognition tasks~\cite{gkikas_rojas_painformer_2025, gkikas_tachos_2024, gkikas_phd_thesis_2025, gkikas_arzate_eeite_pain_2026}, where the same trade-off between model complexity and deployment efficiency recurs.

\IEEEpubidadjcol

Approaches for assessing cognitive load are typically classified as subjective or objective \cite{rojas_debie_2020}. Subjective methods rely on self-reports, with the NASA Task Load Index (NASA-TLX) being the most widely used and thoroughly validated measure of perceived workload \cite{galy_paxion_2018}. These methods are simple to implement and straightforward to interpret; however, they can only be applied after the task is completed and therefore cannot enable continuous monitoring. Objective methods, by contrast, use physiological signals recorded through sensors \cite{hirachan2022measuring}. They are grounded in the well-established link between mental effort and autonomic nervous system activity, which enables direct, real-time estimation of cognitive load \cite{charles_nixon_2019}.

The present study followed a controlled experimental protocol in which 11 participants performed three cognitive tasks: a fluid intelligence reasoning task, a numerical processing task, and an interactive video game. Each task was presented at two difficulty levels (easy and hard) to elicit different levels of cognitive demand.
Continuous electroencephalography (EEG) recordings were collected throughout task performance to capture neural responses across both workload levels.
In addition, a lightweight Transformer-based model is proposed. It is designed for high efficiency and low computational cost, supporting potentially real-time cognitive load estimation in settings that require rapid system adaptation and user support.

\section{Related Work}
Recent advances in cognitive workload assessment increasingly are based on deep learning methods to capture complex neural patterns from neurophysiological signals. EEG-based approaches frequently use CNN and LSTM architectures to learn useful representations from raw or processed data, enabling workload assessment in realistic settings such as augmented reality and multitasking \cite{qin_bulbul_2023, mathews2024eeg}. More specialized network designs further improve EEG-based estimation \cite{siddhad_roy_2024, gkikas_tsiknakis_embc, gkikas_tsiknakis_painvit_2024, gkikas_tsiknakis_thermal_2024, gkikas_chatzaki_2022, gkikas_chatzaki_2023, gkikas_kyprakis_eda_2025}, while recent reviews report that deep learning often outperforms traditional feature-based approaches across many reported scenarios \cite{khingphai_moshfeghi_2025, gkikas_tsiknakis_slr_2023}.
Beyond EEG, functional near-infrared spectroscopy (fNIRS) and multimodal EEG--fNIRS systems have also shown strong results in workload classification. 
Convolutional networks applied to prefrontal fNIRS data can accurately detect workload levels in N-back tasks \cite{park_lee_2023, bargshady_aziz_2025}, while hybrid CNN--LSTM models further improve performance by capturing both spatial and temporal features \cite{khan_asadi_2025}.
In addition, studies have shown that multimodal fusion of EEG and fNIRS signals surpasses single-modality \cite{debie_rohas_2021, saadati_nelson_2020, arif_wang_2024, bunterngchit_wang_2024, li_zhu_2025, deligani_borgheai_2021, farmani_bargshady_2025, khan_chetty_2026}. 
However, the additional hardware complexity and synchronization requirements of multimodal systems limit their practicality outside controlled laboratory environments. EEG remains a compelling modality for real-world deployment, given its temporal resolution, portability, and relatively low acquisition cost --- which explains the continued interest in compact, efficient EEG-based architectures~\cite{gkikas_tiny_2025}.

\section{Methodology}
\begin{figure*}
\begin{center}
\includegraphics[scale=0.20]{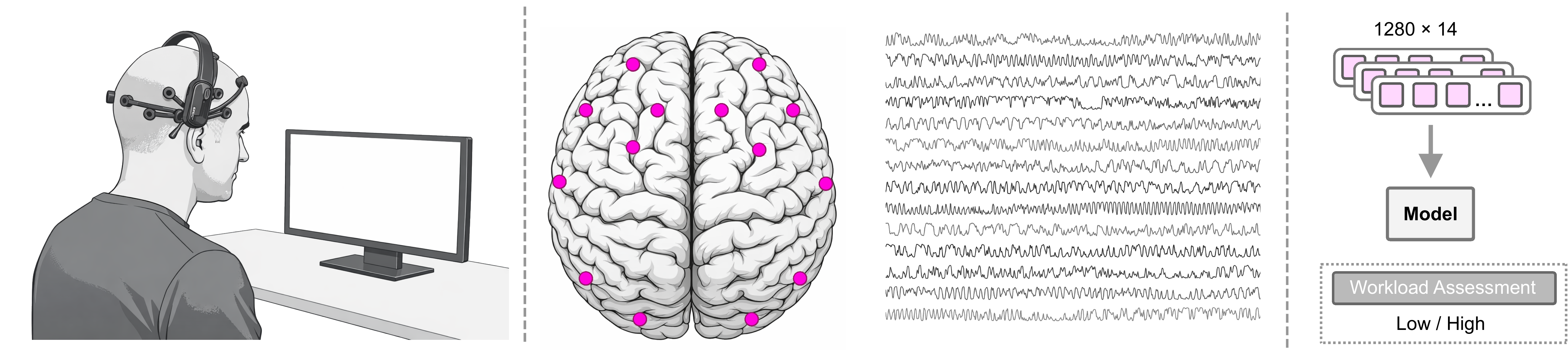} 
\end{center}
\caption{Overview of the study pipeline: (left) EEG data acquisition with the \textit{EMOTIV EPOC} during task performance, (middle) sensor locations over the corresponding cortical regions with the recorded $14$ EEG channels, and (right) the resulting EEG signals structured as $1280 \times 14$ tensor and processed by the proposed model for cognitive workload assessment.}
\label{overview}
\end{figure*}

This section describes the EEG data acquisition process, the corresponding preprocessing pipeline, and the architecture of the proposed \textit{1BT} model used for cognitive workload assessment. Figure \ref{overview} provides an overview of the study.

\subsection{Data Collection}
\label{sec:data_collection}

A pilot study involved eleven participants (mean age $25 \pm 5.5$). None reported neurological disorders or the use of substances affecting the nervous system, such as alcohol or nicotine, on the day of the experiment. All participants provided written informed consent, and the Institutional Review Board approved the study protocol.
EEG data were acquired using the \textit{EMOTIV EPOC}, a wireless neural recording device with 14 measurement channels. Following the international $10$--$20$ system, the recorded channels were AF3, F7, F3, FC5, T7, P7, O1, O2, P8, T8, FC6, F4, F8, and AF4. The sampling frequency was set to $128$~Hz.

Participants completed three activities, and each one was presented at two levels of cognitive difficulty. The activities consisted of a numerical task that required mathematical problem-solving, Raven's Progressive Matrices for abstract reasoning, and the open-source game Flappy Bird, in which participants steered a bird through gaps between pipes. Before starting each activity, they received instructions explaining the procedure.
To reduce order effects, the researchers arranged both the task sequence and the difficulty levels using a strength-3 orthogonal array. All activities appeared on a screen while EEG signals were recorded simultaneously. The experimental interface, together with the task implementations and instruction screens, was developed in Python using standard libraries and extensions.
The dataset is balanced, with each participant contributing $12$ samples per difficulty level across the three tasks, resulting in a total of $792$ samples.

\subsection{Pre-processing}
The EEG signals were preprocessed using the EEGLAB toolbox in MATLAB, with emphasis on removing eye-blink artifacts. Independent Component Analysis (ICA) was applied to decompose the signals and identify components dominated by noise. Components associated with non-neural activity, including ocular and muscular sources, were visually inspected and excluded from the dataset, resulting in cleaner brain-related EEG signals for subsequent analysis.

\subsection{Architecture}
\label{sec:architecture}

\begin{figure}
\begin{center}
\includegraphics[scale=0.25]{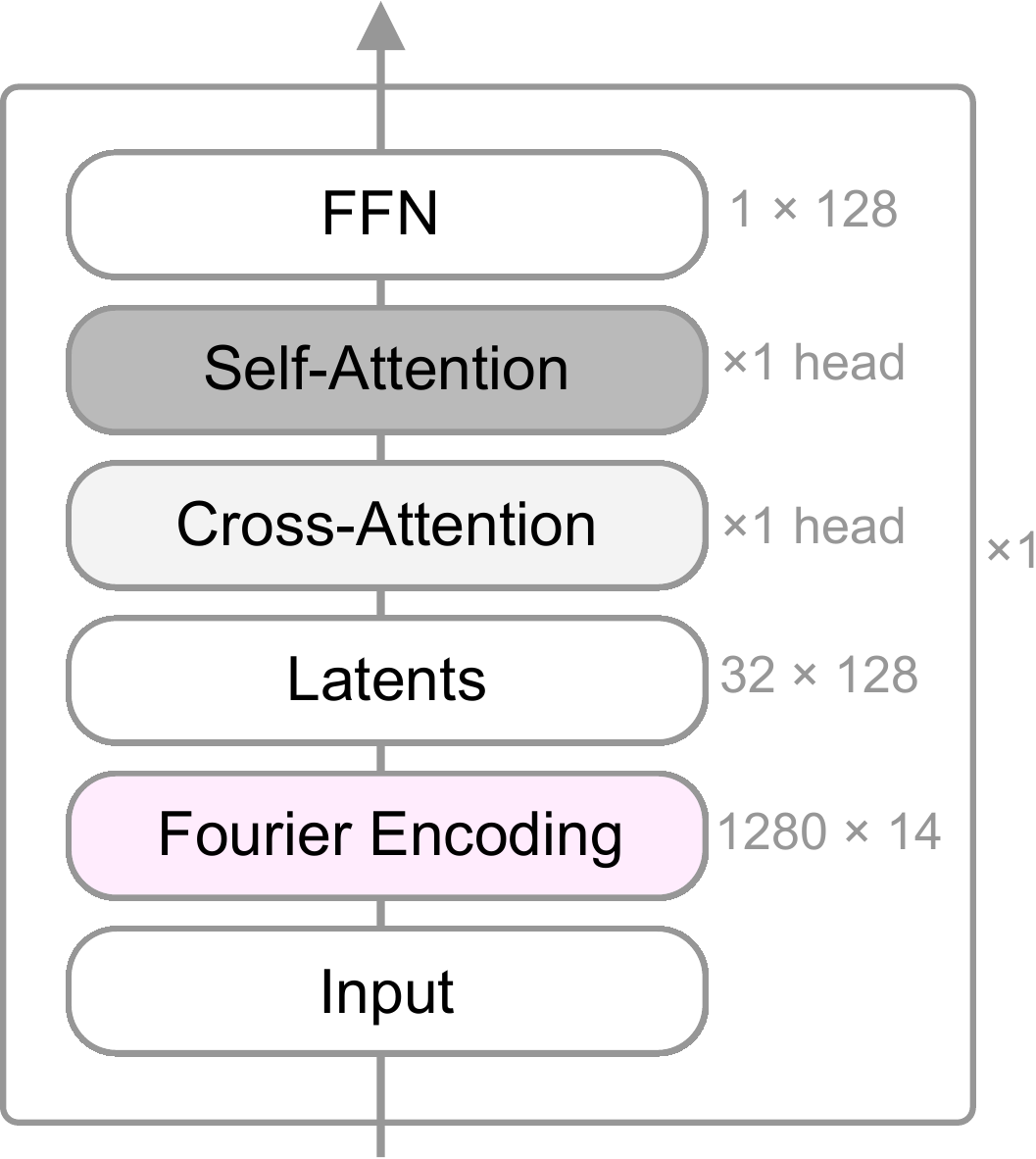} 
\end{center}
\caption{Overview of the proposed transformer-based model and its main architectural components.}
\label{model}
\end{figure}

The proposed architecture directly utilizes stacked multi-channel EEG signals. Each sample is formed by stacking the EEG channels over time, producing a matrix $\mathbf{X} \in \mathbb{R}^{L \times C}$, where $L$ denotes the temporal length and $C$ the number of EEG channels. Since the data are purely temporal, the formulation uses a single input axis ($D=1$). The temporal dimension is reshaped into a sequence of $N=L$ tokens, with each token containing the channel values at the corresponding time index.\\
Positional information is incorporated using Fourier feature encoding. For temporal positions $p \in [-1,1]$, with $K$ frequency components $s_k$ spanning $[1, f_{\max}/2]$ linearly, the encoding is defined as:
\begin{equation}
\gamma(p) = \bigl[\sin(\pi s_1 p), \cos(\pi s_1 p), \ldots, \sin(\pi s_K p), \cos(\pi s_K p), p\bigr],
\end{equation}
where $f_{\max}$ is the maximum frequency. The positional encoding contributes $2K+1$ additional features per temporal position.
The positional features are concatenated with the EEG channel values, creating the token matrix:
\begin{equation}
\mathbf{T} \in \mathbb{R}^{N \times C'}, 
\qquad
C' = C + (2K+1).
\end{equation}
The architecture uses a learnable latent representation as a compact intermediate representation for the full temporal sequence. Specifically, a set of latent vectors, $\mathbf{L} \in \mathbb{R}^{M \times d}$, interacts with the token sequence through a cross-attention mechanism. This allows the model to learn from all time steps and to encode them in a fixed-dimensional latent space.
Cross-attention has previously been shown to serve as an effective bottleneck for biosignal analysis, compressing multi-window temporal representations into compact latent spaces~\cite{gkikas_kyprakis_resp_2025}:
\begin{equation}
\mathbf{L}^{(\ell)}
=
\mathbf{L}^{(\ell-1)}
+
\mathrm{Attn}\!\big(\mathbf{L}^{(\ell-1)},\ \mathbf{T},\ \mathbf{T}\big).
\end{equation}
Multi-head attention is computed as:
\begin{equation}
\mathrm{Attn}(\mathbf{Q},\mathbf{K},\mathbf{V})
=
\mathrm{softmax}\!\left(
\frac{\mathbf{Q}\mathbf{K}^\top}{\sqrt{d_h}}
\right)\mathbf{V},
\end{equation}
with linear projections applied to calculate the queries, keys, and values.
This operation compresses the full token sequence into $M$ latent vectors. In this way, we avoid the self-attention over the entire temporal dimension. The latent states are then updated through self-attention within the latent space:
\begin{equation}
\mathbf{L}^{(\ell)}
\leftarrow
\mathbf{L}^{(\ell)}
+
\mathrm{Attn}\!\big(\mathbf{L}^{(\ell)},\ \mathbf{L}^{(\ell)},\ \mathbf{L}^{(\ell)}\big),
\end{equation}
applied multiple times per layer to enable global interactions among latent vectors. The model uses a single cross-attention layer, and repeated latent self-attention blocks iteratively process the representation.
After the final refinement stage, the latent states are aggregated to form a global embedding of the EEG sequence, which is passed through a linear classification head to produce the output prediction. Figure \ref{model} illustrates the main components of the proposed architecture.


\section{Experimental Evaluation \& Results}
This section presents experiments evaluating EEG signals for cognitive workload assessment using the leave-one-subject-out (LOSO) validation protocol. One of the primary objectives is to identify the most compact and efficient configuration of the proposed architecture. 
We explored the relationship between computational cost and performance by systematically reducing the number of components, model dimensions, and the number of latent vectors.
The corresponding tables report results across three cognitive tasks: IQ (abstract reasoning), MATH (numerical problem-solving), and GAME (video game task), as described in Section \ref{sec:data_collection}. The reported results are organized by task type, and the binary classification labels correspond to the two difficulty levels: easy and hard.

The number of sequential self-attention blocks following each cross-attention module was varied first, spanning configurations from $8$ down to $1$ (Table~\ref{table:table_1}). Reducing depth not only substantially lowers computational cost --- from $3.91$M parameters and $0.26$~GFLOPs at $8$ blocks to $0.69$M and $0.05$~GFLOPs at a single block --- but the single-block configuration also yields the best mean score of $68.37\%$ across all three tasks. Across the tasks, the GAME consistently achieves the highest scores (above $71\%$), while IQ and MATH show lower and more variable results. This reflects the different cognitive dynamics and signal characteristics. We note that the single-block configuration is used as the basis for all subsequent experiments.
The number of attention heads was varied next, reducing from $6$ down to $1$ within the single-block configuration (Table~\ref{table:table_2}). The single-head setup achieves the best mean score of $69.37\%$, with marginal reductions in parameters ($0.46$M) and GFLOPs ($0.04$), which suggests that a simpler self-attention mechanism is sufficient here. 
Performance with $4$ and $2$ heads was almost identical ($68.44\%$ and $68.41\%$), while the single-head model performed slightly better, avoiding the expected decline. 

The combined effect of reducing both the number of latent vectors and their dimensionality is presented in Table~\ref{table:table_3}. Three configurations evaluated against the $32$-latent, $128$-dimensional baseline. By halving the number of latents to $16$ while keeping the $128$-dimensional space, we achieve a mean score of $69.37\%$ --- matching the best result from Table~\ref{table:table_2} --- and simultaneously halve GFLOPs from $0.04$ to $0.02$, consistent with the quadratic scaling of self-attention with respect to latent count. By reducing the latent dimension to $64$ (at either $32$ or $16$ latents), we observe that it degrades performance to approximately $67\%$--$68\%$, which indicates that the representational width is the more important factor relative to the number of latent slots. The $16$-latent, $128$-dimensional setting is therefore the most compact configuration that retains the performance level of the larger baselines, at $0.45$M parameters and $0.02$~GFLOPs.

Head dimensionality within the cross-attention and self-attention modules was assessed last, using the $16$-latent, $128$-dimensional setup from Table~\ref{table:table_3} as the reference (Table~\ref{table:table_4}). All three reduced configurations --- $(64, 32)$, $(32, 64)$, and $(32, 32)$ for cross and self-attention head dimensions respectively --- fall below the $64/64$ baseline, with mean scores of $66.71\%$, $65.92\%$, and $67.38\%$. By reducing the self-attention head dimension from $64$ to $32$, we observe that hurting than reducing the cross-attention head dimension by the same amount. This may indicate that the latent processing step is more sensitive to limited model capacity than the input-to-latent projection. When both dimensions are set to $32$, the loss decreases, raising performance to $67.38\%$, although it remains below the full $64/64$ setting. The computational savings involved are too small to justify the drop, and the $64/64$ configuration is kept as the final architecture.


\begin{table*}
\scriptsize
\caption{Performance comparison for different numbers of self-attention blocks per cross-attention module across three tasks.}
\label{table:table_1}
\begin{center}
\begin{threeparttable}
\begin{tabular}{P{0.8cm}   P{0.70cm}P{0.9cm}P{1.00cm}P{0.90cm}P{0.90cm}P{0.90cm}P{1.40cm}         P{0.9cm} P{1.0cm} P{0.9cm} P{0.9cm} P{0.9cm}}
\toprule

\multirow[c]{4}{*}{Task} &
\multicolumn{7}{c}{Model Configuration} &
\multicolumn{2}{c}{Computational Cost} &
\multicolumn{3}{c}{Performance} \\

\cmidrule(lr){2-8}\cmidrule(lr){9-10}\cmidrule(lr){11-13}
& \shortstack{\#Latents}
& \shortstack{Latent dim}
& \shortstack{\#Cross-attn\\heads}
& \shortstack{\#Self-attn\\heads}
& \shortstack{Cross\\head dim}
& \shortstack{Self-attn\\head dim}
& \shortstack{Self-attn\\per cross-attn}
& \shortstack{Params(M)}
& \shortstack{GFLOPs}
& \shortstack{Accuracy}
& \shortstack{Precision}
& \shortstack{F1} \\

\midrule
\midrule

IQ   &32 &128 &1 &8 &64 &64 &8 &3.91 & 0.26    &66.29{\fontsize{5}{6}\selectfont\,\textpm 9.48} &66.57{\fontsize{5}{6}\selectfont\,\textpm 15.15 } &63.38{\fontsize{5}{6}\selectfont\,\textpm 13.62}\\\hdashline
MATH &32 &128 &1 &8 &64 &64 &8 &3.91 & 0.26    &64.02{\fontsize{5}{6}\selectfont\,\textpm 5.97} &72.22{\fontsize{5}{6}\selectfont\,\textpm 6.08} &59.90{\fontsize{5}{6}\selectfont\,\textpm 8.95}\\\hdashline
GAME &32 &128 &1 &8 &64 &64 &8 &3.91 & 0.26    &71.59{\fontsize{5}{6}\selectfont\,\textpm 7.91} &74.41{\fontsize{5}{6}\selectfont\,\textpm 6.96} &70.33{\fontsize{5}{6}\selectfont\,\textpm 9.21}\\

\rowcolor{gray!15}
\textit{mean} & & & & & & & & & & & \textit{67.64} & \cellcolor{gray!15}\strut \\
\midrule

IQ   &32 &128 &1 &8 &64 &64 &6 &2.99 &0.20    &64.02{\fontsize{5}{6}\selectfont\,\textpm 5.97} &67.48{\fontsize{5}{6}\selectfont\,\textpm 5.35} &61.86{\fontsize{5}{6}\selectfont\,\textpm 8.65}\\\hdashline
MATH &32 &128 &1 &8 &64 &64 &6 &2.99 &0.20    &61.74{\fontsize{5}{6}\selectfont\,\textpm 6.84} &63.12{\fontsize{5}{6}\selectfont\,\textpm 13.55} &57.97{\fontsize{5}{6}\selectfont\,\textpm 11.61}\\\hdashline
GAME &32 &128 &1 &8 &64 &64 &6 &2.99 &0.20    &66.67{\fontsize{5}{6}\selectfont\,\textpm 10.36} &69.06{\fontsize{5}{6}\selectfont\,\textpm 10.20} &65.41{\fontsize{5}{6}\selectfont\,\textpm 11.43}\\

\rowcolor{gray!15}
\textit{mean} & & & & & & & & & & & \textit{64.13} & \cellcolor{gray!15}\strut \\
\midrule

IQ   &32 &128 &1 &8 &64 &64 &4 &2.07 &0.14    &63.64{\fontsize{5}{6}\selectfont\,\textpm 6.90} &65.21{\fontsize{5}{6}\selectfont\,\textpm 14.01} &60.17{\fontsize{5}{6}\selectfont\,\textpm 10.79}\\\hdashline
MATH &32 &128 &1 &8 &64 &64 &4 &2.07 &0.14    &62.88{\fontsize{5}{6}\selectfont\,\textpm 6.01} &67.76{\fontsize{5}{6}\selectfont\,\textpm 7.40} &60.27{\fontsize{5}{6}\selectfont\,\textpm 8.29}\\\hdashline
GAME &32 &128 &1 &8 &64 &64 &4 &2.07 &0.14    &71.21{\fontsize{5}{6}\selectfont\,\textpm 9.31} &70.36{\fontsize{5}{6}\selectfont\,\textpm 15.44} &69.14{\fontsize{5}{6}\selectfont\,\textpm 13.36}\\

\rowcolor{gray!15}
\textit{mean} & & & & & & & & & & & \textit{65.63} & \cellcolor{gray!15}\strut \\
\midrule

IQ   &32 &128 &1 &8 &64 &64 &2 &1.15 &0.08    &66.67{\fontsize{5}{6}\selectfont\,\textpm 9.73} &70.37{\fontsize{5}{6}\selectfont\,\textpm 8.24} &64.32{\fontsize{5}{6}\selectfont\,\textpm 12.45}\\\hdashline
MATH &32 &128 &1 &8 &64 &64 &2 &1.15 &0.08    &65.15{\fontsize{5}{6}\selectfont\,\textpm 4.79} &70.49{\fontsize{5}{6}\selectfont\,\textpm 6.07} &62.72{\fontsize{5}{6}\selectfont\,\textpm 6.44}\\\hdashline
GAME &32 &128 &1 &8 &64 &64 &2 &1.15 &0.08    &69.32{\fontsize{5}{6}\selectfont\,\textpm 8.93} &70.63{\fontsize{5}{6}\selectfont\,\textpm 9.38} &68.82{\fontsize{5}{6}\selectfont\,\textpm 9.13}\\

\rowcolor{gray!15}
\textit{mean} & & & & & & & & & & & \textit{67.61} & \cellcolor{gray!15}\strut \\
\midrule

IQ   &32 &128 &1 &8 &64 &64 &1 &0.69 &0.05    &65.91{\fontsize{5}{6}\selectfont\,\textpm 7.50} &69.40{\fontsize{5}{6}\selectfont\,\textpm 9.24} &64.56{\fontsize{5}{6}\selectfont\,\textpm 7.81}\\\hdashline
MATH &32 &128 &1 &8 &64 &64 &1 &0.69 &0.05    &65.91{\fontsize{5}{6}\selectfont\,\textpm 8.49} &70.34{\fontsize{5}{6}\selectfont\,\textpm 7.75} &63.36{\fontsize{5}{6}\selectfont\,\textpm 11.24}\\\hdashline
GAME &32 &128 &1 &8 &64 &64 &1 &0.69 &0.05    &71.21{\fontsize{5}{6}\selectfont\,\textpm 8.79} &74.71{\fontsize{5}{6}\selectfont\,\textpm 8.38} &69.90{\fontsize{5}{6}\selectfont\,\textpm 9.79}\\

\rowcolor{gray!15}
\textit{mean} & & & & & & & & & & & \textit{\textbf{68.37}} & \cellcolor{gray!15}\strut \\

\bottomrule
\end{tabular}

\begin{tablenotes}[para,flushleft]
\scriptsize
\textit{mean}: the average performance across the three tasks (IQ, MATH, GAME) and the three evaluation metrics (Accuracy, Precision, F1) \space Self-attn per cross-attn: the number of sequential self-attention modules after the cross-attention module 
\end{tablenotes}
\end{threeparttable}
\end{center}
\end{table*}


\begin{table*}
\scriptsize
\caption{Performance comparison for different numbers of self-attention heads across three tasks.}
\label{table:table_2}
\begin{center}
\begin{threeparttable}
\begin{tabular}{P{0.8cm}   P{0.70cm}P{0.9cm}P{1.00cm}P{0.90cm}P{0.90cm}P{0.90cm}P{1.40cm}         P{0.9cm} P{1.0cm} P{0.9cm} P{0.9cm} P{0.9cm}}
\toprule

\multirow[c]{4}{*}{Task} &
\multicolumn{7}{c}{Model Configuration} &
\multicolumn{2}{c}{Computational Cost} &
\multicolumn{3}{c}{Performance} \\

\cmidrule(lr){2-8}\cmidrule(lr){9-10}\cmidrule(lr){11-13}
& \shortstack{\#Latents}
& \shortstack{Latent dim}
& \shortstack{\#Cross-attn\\heads}
& \shortstack{\#Self-attn\\heads}
& \shortstack{Cross\\head dim}
& \shortstack{Self-attn\\head dim}
& \shortstack{Self-attn\\per cross-attn}
& \shortstack{Params(M)}
& \shortstack{GFLOPs}
& \shortstack{Accuracy}
& \shortstack{Precision}
& \shortstack{F1} \\

\midrule
\midrule

IQ   &32 &128 &1 &6 &64 &64 &1 &0.62 &0.05    &67.42{\fontsize{5}{6}\selectfont\,\textpm 4.97} &71.27{\fontsize{5}{6}\selectfont\,\textpm 5.96} &65.94{\fontsize{5}{6}\selectfont\,\textpm 6.58}\\\hdashline
MATH &32 &128 &1 &6 &64 &64 &1 &0.62 &0.05    &64.77{\fontsize{5}{6}\selectfont\,\textpm 7.19} &68.19{\fontsize{5}{6}\selectfont\,\textpm 6.76} &62.64{\fontsize{5}{6}\selectfont\,\textpm 9.77}\\\hdashline
GAME &32 &128 &1 &6 &64 &64 &1 &0.62 &0.05    &67.05{\fontsize{5}{6}\selectfont\,\textpm 9.31} &70.48{\fontsize{5}{6}\selectfont\,\textpm 9.41} &65.40{\fontsize{5}{6}\selectfont\,\textpm 11.28}\\

\rowcolor{gray!15}
\textit{mean} & & & & & & & & & & & \textit{67.02} & \cellcolor{gray!15}\strut \\
\midrule

IQ   &32 &128 &1 &4 &64 &64 &1 &0.55 &0.04    &65.15{\fontsize{5}{6}\selectfont\,\textpm 8.38} &64.82{\fontsize{5}{6}\selectfont\,\textpm 14.52} &62.56{\fontsize{5}{6}\selectfont\,\textpm 11.97}\\\hdashline
MATH &32 &128 &1 &4 &64 &64 &1 &0.55 &0.04    &67.80{\fontsize{5}{6}\selectfont\,\textpm 5.36} &70.69{\fontsize{5}{6}\selectfont\,\textpm 5.57} &66.62{\fontsize{5}{6}\selectfont\,\textpm 6.34}\\\hdashline
GAME &32 &128 &1 &4 &64 &64 &1 &0.55 &0.04    &72.35{\fontsize{5}{6}\selectfont\,\textpm 10.70} &74.66{\fontsize{5}{6}\selectfont\,\textpm 10.15} &71.30{\fontsize{5}{6}\selectfont\,\textpm 11.51}\\

\rowcolor{gray!15}
\textit{mean} & & & & & & & & & & & \textit{68.44} & \cellcolor{gray!15}\strut \\
\midrule

IQ   &32 &128 &1 &2 &64 &64 &1 &0.49 &0.04    &67.05{\fontsize{5}{6}\selectfont\,\textpm 9.14} &73.97{\fontsize{5}{6}\selectfont\,\textpm 6.84} &63.67{\fontsize{5}{6}\selectfont\,\textpm 12.32}\\\hdashline
MATH &32 &128 &1 &2 &64 &64 &1 &0.49 &0.04    &68.18{\fontsize{5}{6}\selectfont\,\textpm 5.69} &71.91{\fontsize{5}{6}\selectfont\,\textpm 4.69} &66.46{\fontsize{5}{6}\selectfont\,\textpm 7.91}\\\hdashline
GAME &32 &128 &1 &2 &64 &64 &1 &0.49 &0.04    &67.05{\fontsize{5}{6}\selectfont\,\textpm 8.79} &73.32{\fontsize{5}{6}\selectfont\,\textpm 6.74} &64.03{\fontsize{5}{6}\selectfont\,\textpm 12.40}\\

\rowcolor{gray!15}
\textit{mean} & & & & & & & & & & & \textit{68.41} & \cellcolor{gray!15}\strut \\
\midrule

IQ   &32 &128 &1 &1 &64 &64 &1 &0.46 &0.04    &68.94{\fontsize{5}{6}\selectfont\,\textpm 11.00} &72.88{\fontsize{5}{6}\selectfont\,\textpm 9.78} &66.83{\fontsize{5}{6}\selectfont\,\textpm 13.39}\\\hdashline
MATH &32 &128 &1 &1 &64 &64 &1 &0.46 &0.04    &67.42{\fontsize{5}{6}\selectfont\,\textpm 6.36} &72.17{\fontsize{5}{6}\selectfont\,\textpm 6.82} &65.56{\fontsize{5}{6}\selectfont\,\textpm 7.56}\\\hdashline
GAME &32 &128 &1 &1 &64 &64 &1 &0.46 &0.04    &68.56{\fontsize{5}{6}\selectfont\,\textpm 9.63} &76.77{\fontsize{5}{6}\selectfont\,\textpm 7.44} &65.21{\fontsize{5}{6}\selectfont\,\textpm 12.45}\\

\rowcolor{gray!15}
\textit{mean} & & & & & & & & & & & \textit{\textbf{69.37}} & \cellcolor{gray!15}\strut \\

\bottomrule
\end{tabular}

\begin{tablenotes}[para,flushleft]
\scriptsize
\end{tablenotes}
\end{threeparttable}
\end{center}
\end{table*}


\begin{table*}
\scriptsize
\caption{Performance comparison for different latent-space configurations across three tasks.}
\label{table:table_3}
\begin{center}
\begin{threeparttable}
\begin{tabular}{P{0.8cm}   P{0.70cm}P{0.9cm}P{1.00cm}P{0.90cm}P{0.90cm}P{0.90cm}P{1.40cm}         P{0.9cm} P{1.0cm} P{0.9cm} P{0.9cm} P{0.9cm}}
\toprule

\multirow[c]{4}{*}{Task} &
\multicolumn{7}{c}{Model Configuration} &
\multicolumn{2}{c}{Computational Cost} &
\multicolumn{3}{c}{Performance} \\

\cmidrule(lr){2-8}\cmidrule(lr){9-10}\cmidrule(lr){11-13}
& \shortstack{\#Latents}
& \shortstack{Latent dim}
& \shortstack{\#Cross-attn\\heads}
& \shortstack{\#Self-attn\\heads}
& \shortstack{Cross\\head dim}
& \shortstack{Self-attn\\head dim}
& \shortstack{Self-attn\\per cross-attn}
& \shortstack{Params(M)}
& \shortstack{GFLOPs}
& \shortstack{Accuracy}
& \shortstack{Precision}
& \shortstack{F1} \\

\midrule
\midrule

IQ   &32 &64  &1 &1 &64 &64 &1 &0.13 &0.02    &68.94{\fontsize{5}{6}\selectfont\,\textpm 9.12} &74.41{\fontsize{5}{6}\selectfont\,\textpm 7.26} &66.47{\fontsize{5}{6}\selectfont\,\textpm 12.18}\\\hdashline
MATH &32 &64  &1 &1 &64 &64 &1 &0.13 &0.02    &65.53{\fontsize{5}{6}\selectfont\,\textpm 7.96} &70.46{\fontsize{5}{6}\selectfont\,\textpm 7.61} &62.67{\fontsize{5}{6}\selectfont\,\textpm 10.44}\\\hdashline
GAME &32 &64  &1 &1 &64 &64 &1 &0.13 &0.02    &64.39{\fontsize{5}{6}\selectfont\,\textpm 7.82} &68.71{\fontsize{5}{6}\selectfont\,\textpm 6.66} &61.73{\fontsize{5}{6}\selectfont\,\textpm 10.46}\\

\rowcolor{gray!15}
\textit{mean} & & & & & & & & & & & \textit{67.03} & \cellcolor{gray!15}\strut \\
\midrule

IQ   &16 &128 &1 &1 &64 &64 &1 &0.45 &0.02    &69.32{\fontsize{5}{6}\selectfont\,\textpm 7.80} &71.22{\fontsize{5}{6}\selectfont\,\textpm 8.57} &68.55{\fontsize{5}{6}\selectfont\,\textpm 8.20}\\\hdashline
MATH &16 &128 &1 &1 &64 &64 &1 &0.45 &0.02    &67.05{\fontsize{5}{6}\selectfont\,\textpm 7.63} &72.73{\fontsize{5}{6}\selectfont\,\textpm 7.28} &64.43{\fontsize{5}{6}\selectfont\,\textpm 10.21}\\\hdashline
GAME &16 &128 &1 &1 &64 &64 &1 &0.45 &0.02    &69.70{\fontsize{5}{6}\selectfont\,\textpm 8.90} &73.65{\fontsize{5}{6}\selectfont\,\textpm 6.72} &67.70{\fontsize{5}{6}\selectfont\,\textpm 11.59}\\

\rowcolor{gray!15}
\textit{mean} & & & & & & & & & & & \textit{\textbf{69.37}} & \cellcolor{gray!15}\strut \\
\midrule

IQ   &16 &64  &1 &1 &64 &64 &1 &0.13 &0.01    &70.08{\fontsize{5}{6}\selectfont\,\textpm 7.71} &72.17{\fontsize{5}{6}\selectfont\,\textpm 7.55} &68.88{\fontsize{5}{6}\selectfont\,\textpm 9.29}\\\hdashline
MATH &16 &64  &1 &1 &64 &64 &1 &0.13 &0.01    &63.64{\fontsize{5}{6}\selectfont\,\textpm 5.92} &67.35{\fontsize{5}{6}\selectfont\,\textpm 7.87} &61.95{\fontsize{5}{6}\selectfont\,\textpm 7.02}\\\hdashline
GAME &16 &64  &1 &1 &64 &64 &1 &0.13 &0.01    &68.94{\fontsize{5}{6}\selectfont\,\textpm 7.61} &70.14{\fontsize{5}{6}\selectfont\,\textpm 7.19} &68.27{\fontsize{5}{6}\selectfont\,\textpm 8.16}\\

\rowcolor{gray!15}
\textit{mean} & & & & & & & & & & & \textit{67.94} & \cellcolor{gray!15}\strut \\

\bottomrule
\end{tabular}

\begin{tablenotes}[para,flushleft]
\scriptsize
\end{tablenotes}
\end{threeparttable}
\end{center}
\end{table*}


\begin{table*}
\scriptsize
\caption{Performance comparison for different attention head-dimension configurations across three tasks.}
\label{table:table_4}
\begin{center}
\begin{threeparttable}
\begin{tabular}{P{0.8cm}   P{0.70cm}P{0.9cm}P{1.00cm}P{0.90cm}P{0.90cm}P{0.90cm}P{1.40cm}         P{0.9cm} P{1.0cm} P{0.9cm} P{0.9cm} P{0.9cm}}
\toprule

\multirow[c]{4}{*}{Task} &
\multicolumn{7}{c}{Model Configuration} &
\multicolumn{2}{c}{Computational Cost} &
\multicolumn{3}{c}{Performance} \\

\cmidrule(lr){2-8}\cmidrule(lr){9-10}\cmidrule(lr){11-13}
& \shortstack{\#Latents}
& \shortstack{Latent dim}
& \shortstack{\#Cross-attn\\heads}
& \shortstack{\#Self-attn\\heads}
& \shortstack{Cross\\head dim}
& \shortstack{Self-attn\\head dim}
& \shortstack{Self-attn\\per cross-attn}
& \shortstack{Params(M)}
& \shortstack{GFLOPs}
& \shortstack{Accuracy}
& \shortstack{Precision}
& \shortstack{F1} \\

\midrule
\midrule

IQ   &16 &128 &1 &1 &64 &32 &1 &0.44 &0.02    &68.18{\fontsize{5}{6}\selectfont\,\textpm 6.22} &71.49{\fontsize{5}{6}\selectfont\,\textpm 7.32} &66.95{\fontsize{5}{6}\selectfont\,\textpm 6.81}\\\hdashline
MATH &16 &128 &1 &1 &64 &32 &1 &0.44 &0.02    &64.39{\fontsize{5}{6}\selectfont\,\textpm 7.82} &70.52{\fontsize{5}{6}\selectfont\,\textpm 8.00} &61.36{\fontsize{5}{6}\selectfont\,\textpm 10.65}\\\hdashline
GAME &16 &128 &1 &1 &64 &32 &1 &0.44 &0.02    &65.53{\fontsize{5}{6}\selectfont\,\textpm 8.90} &68.56{\fontsize{5}{6}\selectfont\,\textpm 8.71} &63.43{\fontsize{5}{6}\selectfont\,\textpm 10.69}\\

\rowcolor{gray!15}
\textit{mean} & & & & & & & & & & & \textit{66.71} & \cellcolor{gray!15}\strut \\
\midrule

IQ   &16 &128 &1 &1 &32 &64 &1 &0.44 &0.02    &64.77{\fontsize{5}{6}\selectfont\,\textpm 6.25} &66.09{\fontsize{5}{6}\selectfont\,\textpm 5.72} &63.63{\fontsize{5}{6}\selectfont\,\textpm 7.29}\\\hdashline
MATH &16 &128 &1 &1 &32 &64 &1 &0.44 &0.02    &62.88{\fontsize{5}{6}\selectfont\,\textpm 6.01} &69.66{\fontsize{5}{6}\selectfont\,\textpm 8.48} &59.68{\fontsize{5}{6}\selectfont\,\textpm 8.31}\\\hdashline
GAME &16 &128 &1 &1 &32 &64 &1 &0.44 &0.02    &68.18{\fontsize{5}{6}\selectfont\,\textpm 8.00} &72.18{\fontsize{5}{6}\selectfont\,\textpm 7.13} &66.24{\fontsize{5}{6}\selectfont\,\textpm 10.60}\\

\rowcolor{gray!15}
\textit{mean} & & & & & & & & & & & \textit{65.92} & \cellcolor{gray!15}\strut \\
\midrule

IQ   &16 &128 &1 &1 &32 &32 &1 &0.43 &0.02    &67.05{\fontsize{5}{6}\selectfont\,\textpm 8.23} &71.20{\fontsize{5}{6}\selectfont\,\textpm 8.44} &65.25{\fontsize{5}{6}\selectfont\,\textpm 9.62}\\\hdashline
MATH &16 &128 &1 &1 &32 &32 &1 &0.43 &0.02    &65.15{\fontsize{5}{6}\selectfont\,\textpm 4.08} &68.99{\fontsize{5}{6}\selectfont\,\textpm 5.99} &63.56{\fontsize{5}{6}\selectfont\,\textpm 4.65}\\\hdashline
GAME &16 &128 &1 &1 &32 &32 &1 &0.43 &0.02    &67.42{\fontsize{5}{6}\selectfont\,\textpm 9.86} &73.16{\fontsize{5}{6}\selectfont\,\textpm 8.86} &64.64{\fontsize{5}{6}\selectfont\,\textpm 12.40}\\

\rowcolor{gray!15}
\textit{mean} & & & & & & & & & & & \textit{\textbf{67.38}} & \cellcolor{gray!15}\strut \\

\bottomrule
\end{tabular}

\begin{tablenotes}[para,flushleft]
\scriptsize
\end{tablenotes}
\end{threeparttable}
\end{center}
\end{table*}


\section{Discussion \& Conclusion}

This work introduces \textit{1BT}, a One-Block Transformer for EEG-based cognitive workload assessment. The model encodes the multi-channel EEG signals into a small set of learnable latent vectors using a single cross-attention and a lightweight self-attention layer. A structured ablation across four dimensions (self-attention depth, attention heads, number of latents, and head dimensionality) was used to identify the most efficient configuration.

One important outcome concerns the relationship between the model's size and performance. Reducing self-attention blocks from $8$ to $1$ improved mean performance while cutting parameters from $3.91$M to $0.69$M, and a single attention head outperformed all multi-head variants. The final configuration --- $16$ latent vectors of dimension $128$, one cross-attention head, one self-attention head, and one self-attention block --- achieves a mean score of $69.37\%$ across the IQ, MATH, and GAME tasks at $0.45$M parameters and $0.02$~GFLOPs. Taken together, the results point toward representational width being more important than depth or attention diversity, and suggest that excess capacity introduces redundancy rather than benefit in this setting.

The low computational cost and EEG-only design make \textit{1BT} well-suited for real-time deployment in resource-constrained settings~\cite{antonogiorgakis_britzolakis_2019}. One direct application is human-robot interaction, where cognitive state monitoring can support adaptive and empathetic robot behavior, including gaze-aware and perceptually-grounded systems~\cite{arzate_when_2025, arzate_hri_2025, fang_moleron_2024, kruger_oshima_2026, hessels_fang_2026, vazquez_cruz_gkikas_2026}. Evaluation on larger datasets and experimentation across a broader range of cognitively demanding tasks remain necessary before being applied to real-world settings.

\bibliographystyle{IEEEtran}
\bibliography{library}

\appendix
\section{Appendix}

\subsection{Complementary Experiments}

In this section, we present additional experiments using \textit{EEGNet} \cite{lawhern_solon_2018}.
\textit{EEGNet} is a compact convolutional neural network designed specifically for EEG-based brain-computer interfaces. The model extracts features in two main steps. It first applies temporal convolutions to capture frequency-related temporal patterns, and then uses depthwise spatial convolutions to learn how signals from different electrodes interact. After that, a separable convolution together with strong average pooling reduces the feature representation before the final linear classification layer.
The result is an extremely lightweight model with no attention mechanism, making it a principled and well-established baseline for comparison with transformer-based approaches.
In Table \ref{table:eegnet}, the performances across the three tasks are presented. We use the same training configuration, including augmentation techniques, regularization methods, and schedulers, as utilized throughout this work. We observe that across all tasks and all three reported metrics, \textit{EEGNet} underperforms the proposed model. The mean performance across the three metrics is $64.47\%$, approximately $5\%$ lower than the $69.47\%$ reported in the previous section. At the same time, the computational cost is on par with the proposed model, with $0.01$M parameters and $0.02$ GFLOPs.
Table \ref{table:training} summarizes the training and regularization details used across all presented experiments.

\begin{table}[h]
\scriptsize
\caption{Performances using the \textit{EEGNet} across three tasks.}
\label{table:eegnet}
\begin{center}
\begin{threeparttable}
\begin{tabular}{P{0.8cm} P{0.9cm} P{1.0cm} P{0.9cm} P{0.9cm} P{0.9cm}}
\toprule

\multirow[c]{4}{*}{Task} &
\multicolumn{2}{c}{Computational Cost} &
\multicolumn{3}{c}{Performance} \\

\cmidrule(lr){2-3}\cmidrule(lr){4-6}
& \shortstack{Params(M)}
& \shortstack{GFLOPs}
& \shortstack{Accuracy}
& \shortstack{Precision}
& \shortstack{F1} \\

\midrule
\midrule

IQ   &0.01 &0.02 &65.15{\fontsize{5}{6}\selectfont\,\textpm 6.47} &65.90{\fontsize{5}{6}\selectfont\,\textpm 6.55} &64.75{\fontsize{5}{6}\selectfont\,\textpm 6.67} \\\hdashline
MATH &0.01 &0.02 &63.64{\fontsize{5}{6}\selectfont\,\textpm 13.66} &63.95{\fontsize{5}{6}\selectfont\,\textpm 13.89} &63.32{\fontsize{5}{6}\selectfont\,\textpm 13.83} \\\hdashline
GAME &0.01 &0.02 &64.39{\fontsize{5}{6}\selectfont\,\textpm 9.46} &64.93{\fontsize{5}{6}\selectfont\,\textpm 9.89} &64.19{\fontsize{5}{6}\selectfont\,\textpm 9.41} \\

\rowcolor{gray!15}
\textit{mean} & & & &  \textit{64.47} & \cellcolor{gray!15}\strut \\

\bottomrule
\end{tabular}
\begin{tablenotes}[para,flushleft]
\scriptsize
\end{tablenotes}
\end{threeparttable}
\end{center}
\end{table}

\begin{table}[h]
\scriptsize
\caption{Regularization, and training configuration.}
\label{table:training}
\begin{center}
\begin{tabular}{P{2.5cm} P{4.5cm}}
\toprule
Parameter & Value \\
\midrule
\midrule
\textit{Label smoothing}     & 0.10 \\\hdashline
\textit{Attention dropout}   & 0.10 \\\hdashline
\textit{FF dropout}          & 0.10 \\
\midrule
\midrule
Optimizer                    & AdamW \\\hdashline
Learning rate $\eta$         & $1\times10^{-4}$ \\\hdashline
LR schedule                  & Cosine annealing \\\hdashline
Weight decay $\lambda$       & $0.05$ \\\hdashline
Epochs                       & 200 \\\hdashline
Batch size                   & 32 \\
\bottomrule
\end{tabular}
\end{center}
\end{table}

\end{document}